\documentclass[aps,twocolumn,superscriptaddress]{revtex4-1} 
\usepackage[english]{babel}
\usepackage{amsmath, bm, amssymb}
\usepackage{mathtools}
\usepackage{siunitx}
\usepackage{environ}
\usepackage{mathrsfs}
\usepackage{braket}
\usepackage[]{hyperref}
\DeclareMathAlphabet{\mathscrlower}{OT1}{pzc}{m}{it} 
\usepackage{prettyref}
\usepackage{graphicx}
\usepackage{booktabs}
\usepackage{array}
\usepackage{multirow}
\usepackage{dcolumn}
\usepackage{threeparttable}
\usepackage{threeparttablex}
\usepackage{placeins}
\usepackage[stable]{footmisc}
\usepackage{mhchem}
\usepackage{xfrac}

\newcommand{\pauli}{\boldsymbol{\sigma}}

\newcommand{\diraca}{\vec{\boldsymbol{\alpha}}}

\newcommand{\pos}{\vec{r}}

\newcommand{\Sum}[2]{\sum\limits_{#1}^{#2}}
\newcommand{\parantheses}[1]{\left(#1\right)}
\newcommand{\brackets}[1]{\left[#1\right]}
\newcommand{\braces}[1]{\left\{ #1\right\}}
\let\nablatmp\nabla
\renewcommand{\nabla}{\vec{\nablatmp}}
\DeclarePairedDelimiter\abs{\lvert}{\rvert}
\makeatletter
\let\oldabs\abs
\def\abs{\@ifstar{\oldabs}{\oldabs*}}

\newrefformat{fig}{Figure~\ref{#1}}
\newrefformat{tab}{Table~\ref{#1}}

\AtBeginDocument{
\heavyrulewidth=.08em
\lightrulewidth=.05em
\cmidrulewidth=.03em
\belowrulesep=.65ex
\belowbottomsep=0pt
\aboverulesep=.4ex
\abovetopsep=0pt
\cmidrulesep=\doublerulesep
\cmidrulekern=.5em
\defaultaddspace=.5em
}
\begin{document}
\title{Rigorous extension of semilocal collinear functionals to noncollinear DFT using $SU(2)$ rotations}
\date{\today}
\author{Konstantin Gaul}
\affiliation{Helmholtz Institut Mainz, 55099 Mainz, Germany}
\affiliation{GSI Helmholtzzentrum für Schwerionenforschung GmbH, 64291 Darmstadt, Germany}
\affiliation{Institut f\"ur Physik, Johannes Gutenberg-Universit\"at Mainz, 55099 Mainz, Germany}
\email[]{konstantin.gaul@uni-mainz.de}
\begin{abstract}
In the presence of spin-orbit coupling and in geometrically frustrated
materials, a noncollinear treatment the magnetization density is essential.
However, in density functional theory most exchange--correlation functional approximations were originally
developed for locally collinear magnetization. Many practical approaches to
noncollinear DFT have emerged over the past decade. However, a first-principles
connection between widely used semilocal collinear functionals and their
noncollinear generalizations remains lacking. In this work, a locally exact
relation between collinear and noncollinear exchange--correlation functionals is
derived at the level of gradient expansions within a $u(2)$ matrix
representation of the energy functional. Within this framework, collinear semilocal
variables naturally acquire distinct dependencies on
transverse and longitudinal magnetization gradient components. The widely used
Scalmani--Frisch scheme emerges as a first-order approximation. The transformation
of collinear functional derivatives to noncollinear space is implemented through
numerically robust $SU(2)$ rotations. A consistent description of local
magnetic torques is demonstrated for the prototypical spin-frustrated
Cr$_3$ cluster. The approach further extends to fully nonlocal functionals and provides a
direct route towards numerically stable relativistic response calculations. The influence on
magnetic properties in presence of spin-orbit coupling is illustrated through
calculations of hyperfine couplings in the high-spin ground states of uranium and the uranium
ion.
\end{abstract}

\maketitle

\section{Introduction}
Noncollinear magnetism is essential for a qualitatively correct description
of many material properties, including geometrical frustration, the spin Hall
effect, topological insulating phases
\cite{comaskey:2022,bodo:2022,pu:2023,chen:2022}.  Moreover, noncollinearity
is crucial for describing systems containing heavy elements, in which
spin-orbit coupling breaks the spin rotational symmetry
\cite{eschrig:1999,wullen:2002,wang:2003,bast:2009}. 

Semilocal exchange--correlation functionals are typically constructed under the
assumption of locally collinear magnetization. A first-principles relation
between locally collinear and locally noncollinear functionals within the
local spin-density approximation (LSDA) \cite{barth:1972,rajagopal:1973} was
demonstrated by K\"ubler nearly forty years ago
\cite{kubler:1988,kubler:1988a}. Noncollinear LSDA was successfully applied in
relativistic electronic-structure calculations
\cite{eschrig:1999,wullen:2002,wang:2003,anton:2002,anton:2004}.  In contrast,
noncollinear DFT calculations employing semilocal functionals rely on ad hoc
approaches \cite{peralta:2004,peralta:2007,wullen:2007,scalmani:2012,pu:2022},
dedicated noncollinear functionals \cite{eich:2013}, functionals that
introduce magnetic torques through spin currents
\cite{tancogne-dejean:2022,huebsch:2025}, and orbital-optimized exact-exchange
approaches \cite{sharma:2007,ullrich:2018}.  The most widely used approaches based on
existing collinear functionals are the so-called canonical approach
\cite{peralta:2004,peralta:2007,wullen:2007} and the approach by Scalmani and
Frisch (SF) \cite{scalmani:2012,bulik:2013}.  Although these approaches provide
a powerful framework for many applications, they exhibit several shortcomings,
including vanishing local magnetic torques
\cite{scalmani:2012,bulik:2013,pu:2023}, ill-defined functional derivatives in
low-spin-density regions
\cite{egidi:2017,komorovsky:2019,desmarais:2021,pu:2022}, and an inconsistent
description of the local spin rotations \cite{eich:2013a,pu:2023}.  More
recently, Pu \emph{et al.} proposed an alternative, multicollinear approach
based on a local averaging over all spin orientations
\cite{pu:2023}. Although the approach possesses several advantages, including well-defined
functional derivatives, the correct collinear limit and nonzero local torques,
the approach of Ref.\,\cite{pu:2023} fails to reproduce qualitatively the local
magnetic torque structure observed in exact exchange (EXX-KLI) calculations
\cite{tancogne-dejean:2022}.

In this work, we derive a rigorous local connection between arbitrary collinear and
noncollinear functionals by expressing the exchange--correlation
functional in $u(2)$-matrix representation space. We derive
a first-principles connection between the gradient expansions of the collinear and
noncollinear energy functionals. An energy-invariant
relationship between collinear and noncollinear exchange--correlation
functionals is established. The formalism is connected to previous approaches through first
order perturbation theory. We obtain well-defined and 
numerically robust functional derivatives and provide a straightforward
implementation scheme for transforming collinear functionals and their
derivatives to noncollinear space through local $SU(2)$ rotations. The
approach is applicable to arbitrary approximate collinear functionals, including local,
semilocal, and nonlocal forms. Implications for geometrically frustrated materials
are investigated through the local torque structure of the prototypical \ce{Cr3}
cluster. Applications to relativistic electronic-structure theory are demonstrated through
calculations of hyperfine coupling constants for uranium and its ionized species. 

\section{Theory}
\subsection{$u(2)$ representation of the energy functional}
The noncollinear electronic energy functional 
is typically written as \cite{barth:1972,kubler:1988}
\begin{equation}
E[\rho,\vec{m}] = \int\mathrm{d}^3 r\, F[\rho(\pos),\vec{m}(\pos)]\,,
\end{equation}
where $\rho$ is the particle density and $\vec{m}$ is the magnetization
density. Alternatively, the energy functional may be formulated as a
matrix-valued functional in $u(2)$ representation space. Defining 
$\rho_0=\rho,\,\rho_1=m_x,\,\rho_2=m_y,\,\rho_3=m_z$, we
introduce a matrix-valued density in the $u(2)$ representation space as \cite{eschrig:1999,anton:2004}
\begin{equation}
\bm{\rho}(\pos) = \rho_\mu(\pos) \mathfrak{G}^\mu\,,
\end{equation}
where we employ tensor-index notation: Greek letters indices run over
$\mu,\nu,\dots=0,1,2,3$, Roman letters indices run over $i,j,k,\dots=1,2,3$,
and spin-space indices are denoted by $a,b,c,\dots=\alpha,\beta$.  We use
Einstein summation convention, such that repeated upper and lower indices are
implicitly summed, e.g.\ $v_\mu w^\mu$. We introduce the Hermitian generator basis
$\mathfrak{G}^\mu$, associated with $u(2)$,  in terms of the Pauli spin matrices $\pauli^\mu$ together with
the $2\times2$ identity matrix, defined as $\pauli^0=\bm{1}$, such that
$\mathfrak{G}^\mu=\pauli^\mu/2$. We define a scalar projection functional as
\begin{equation}
\mathcal{P}(\bm{A})=\mathrm{Tr}\brackets{\Sum{\mu=0}{3}\mathfrak{G}^\mu\bm{A}}\,.
\end{equation}
With these definitions, the general noncollinear energy functional may be written as
\begin{equation}
E[\bm{\rho}] = \int\mathrm{d}^3 r\, \underbrace{\mathcal{P}\parantheses{\bm{F}[\bm{\rho}(\pos)]}}_{F[\rho(\pos),\vec{m}(\pos)]}\,
\end{equation}
where $\bm{F}$ is a matrix-valued functional in the $u(2)$ representation space. 

At a each point $\vec{r}$ in real space, noncollinear (ncl) and collinear
(cl) representations of the energy functional are related by a $SU(2)$ rotation $\bm{U}_0(\pos)$ such that 
\begin{equation}
\bm{U}^\dagger_0(\pos)\bm{\rho}(\pos)\bm{U}_0(\pos) = \tilde{\bm{\rho}}(\pos)\,,
\end{equation}
where $\tilde{\bm{\rho}}$ is diagonal with eigenvalues $\rho_+$, $\rho_-$.  The
collinear particle and spin densities are  $\rho=\rho_+(\pos)+\rho_-(\pos)$,
$s=\rho_+(\pos)-\rho_-(\pos)$. Because $\bm{U}_0$ is unitary, the transformation
preserves the energy. The noncollinear energy may therefore be expressed in terms
of a locally collinear functional as
\begin{equation}
E[\bm{\rho}] = \int\mathrm{d}^3 r\, \mathcal{P}_{\mathrm{cl}}\parantheses{\bm{F}_\mathrm{cl}[\tilde{\bm{\rho}}(\pos)]}\,
\end{equation}
where 
\begin{equation}
\bm{F}_\mathrm{cl}[\tilde{\bm{\rho}}(\pos)] = \bm{U}_0^\dagger(\pos) \bm{F}[\bm{\rho}(\pos)] \bm{U}_0(\pos)\,
\end{equation}
and the collinear norm projector is 
\begin{equation}
\mathcal{P}_{\mathrm{cl}}(\bm{A})=\mathrm{Tr}\brackets{\bm{U}_0^\dagger(\pos)\parantheses{\Sum{\mu=0}{3}\mathfrak{G}^\mu}\bm{U}_0(\pos)\bm{A}}\,.
\label{eq: projcoll}
\end{equation}
Eq. \prettyref{eq: projcoll} establishes a formally exact local relation between a
noncollinear functional and its locally collinear representation. 

\subsection{Approximate functionals and noncollinear gradient expansions}
The formulation introduced in the previous section is not of immediately applicable to approximate
energy functionals. Semilocal and nonlocal functionals are commonly constructed from gradient
expansions of the energy functional and involve coefficient tensors connected to rotational
invariants constructed from derivatives of $\bm{\rho}$. In the following, we will
discuss the gradient expansion of the local matrix functional, which may subsequently be
embedded into semilocal or nonlocal approximations.  

Generalizing the gradient-expansion formalism of
Refs.\,\cite{hohenberg:1964,langreth:1980,langreth:1983} to $u(2)$ matrix
representations of the energy functional, we obtain
\begin{equation}
\begin{aligned}
\bm{F}[\bm{\rho}(\pos)]
 & =\bm{F}^{(0)}(\bm{\rho}(\pos))
   + \mathrm{Tr}\brackets{\bm{F}^{(2)}_2(\bm{\rho}(\pos)) \cdot\bm{\gamma}}_2 \\
 & + \mathrm{Tr}\brackets{\bm{F}^{(2)}_4(\bm{\rho}(\pos)) \cdot\left(\Delta\bm{\rho}\otimes\Delta\bm{\rho} \right)}_2 \\
 & + \mathrm{Tr}\brackets{\bm{F}^{(3)}_4(\bm{\rho}(\pos)) \cdot\left( \Delta\bm{\rho}\otimes\bm{\gamma}\right)}_3 \\
 & + \mathrm{Tr}\brackets{\bm{F}^{(4)}_4(\bm{\rho}(\pos)) \cdot\left( \bm{\gamma}\otimes\bm{\gamma} \right)}_4 \\
 & + \mathcal{O}(\partial^6)\,,
\end{aligned}
\label{eq: ggancl}
\end{equation}
where $\bm{F}^{(m)}_k(\bm{\rho}(\pos))$ is a coefficient-tensor in
$u(2)^{\otimes m}$ representation space parametrizing  the
exchange--correlation (XC) functional. $\mathrm{Tr}\brackets{}_m$ denotes the
partial trace over the $m$ innermost tensor indices in $u(2)^{\otimes (n+m)}$
representation space. We introduce a rotationally invariant variable of
derivative order $k$ and density order $m$:
$\bm{\partial}^{(m)}_k=\bigotimes_{i=1}^m\brackets{(\partial)^{k_i}\bm{\rho}}$
with $\sum_i k_i=k$. This notation should not to be confused with a partial
derivative operator. 

Up to order four [see eq.\,\prettyref{eq: ggancl}] only two additional
variables are introduced: the squared gradient tensor
$\bm{\partial}^{(2)}_2=\bm{\gamma}=
\partial_k\bm{\rho}\otimes\partial^k\bm{\rho}=\gamma_{\mu\nu}\bm{\mathfrak{G}}^\mu\otimes\bm{\mathfrak{G}}^\nu$,
with $\gamma_{\mu\nu}=\partial_k\rho_\mu\partial^k\rho_\nu$ and the Laplacian
density tensor $\bm{\partial}^{(1)}_2=\Delta\bm{\rho}=\partial_k\partial^k\bm{\rho}$.
We emphasize that no specific spin-structure is assumed for $F$. Instead, allow tensor components are allowed to be 
distinct and nonzero. 

Energy-invariant transformations of $\bm{F}$ must be unitary. In order to
connect to collinear space the transformation must simultaneously diagonalize
the coefficient tensors and gradient variables in $u(2)^{\otimes m}$
representation space for each term in (\ref{eq: ggancl}). It is important to
note that the exact noncollinear functional can only be represented exactly
within a locally collinear parametrization if all coefficient tensors commute
with the corresponding gradient variables. This is not generally the case. From
the commutation properties of Pauli matrices we find for the first term of the
gradient expansion:
\begin{equation}
\brackets{\bm{F}^{(2)}_2(\bm{\rho}(\pos)),\bm{\gamma}}\overset{!}{=}0\Leftrightarrow
\gamma_{kl} F^{(2)}_{2,\mu mn}
\epsilon^{mk}{}_{i}\epsilon^{nl}_{j}\overset{!}{=}0\forall\mu\,,
\end{equation} 
where $k,l,m,n=1,2,3$ are indices in spin-space. Even if the
commutator does not vanish, the functional may still become representable at
higher order in the gradient expansion. Otherwise, the locally collinear
representation becomes an approximation. In practical applications, additional
approximations entering semilocal collinear functionals are likely to have a
significantly larger effect.

The different variables $\bm{\rho}$, $\bm{\gamma}$ and $\Delta\bm{\rho}$ do not
commute. We therefore must determine the eigenvectors associated with each
variable to formulate a relation to the collinear functional:

\begin{equation}
\begin{aligned}
\bm{U}_{\gamma}^\dagger(\pos)\bm{\gamma}\bm{U}_{\gamma}(\pos)&=\widetilde{\bm{\gamma}}\,, \\
\bm{U}_{\Delta}^\dagger(\pos)\Delta\bm{\rho}\bm{U}_{\Delta}(\pos)&=\widetilde{\Delta\bm{\rho}}\,,
\end{aligned}
\label{eq: variable_transform}
\end{equation}
with analogous relations holding for higher-order terms. 

The collinear gradient expansion is obtained by independently rotating each
variable and its corresponding coefficient tensor from noncollinear
representation space to its diagonal representation. For each coefficient
tensor, we define

\begin{equation}
\widetilde{\bm{F}^{(m)}_k} = [\bm{U}^{(m)}_k]^\dagger\bm{F}^{(m)}_k\bm{U}^{(m)}_k\,.
\end{equation}
Here, the induced $SU(2)$ rotation in the $u(2)^{\otimes m}$ representation space is defined as
\begin{equation}
\bm{U}^{(m)}_{i_1\dots i_m}(\pos) = \bigotimes\limits_{k=1}^{m}
\bm{U}_{i_k}(\pos)\,,
\end{equation} 
where $i_k \in\braces{0,\gamma,\Delta,\dots}$. This transformation defines the
noncollinear functional in terms of a collinear functional through an
energy-invariant representation change. Consequently, each term in the gradient
expansion may be expressed through an $SU(2)$ rotation of the corresponding 
collinear representation:
\begin{equation}
\mathrm{Tr}\brackets{ 
\bm{U}^{(m)}_k\widetilde{\bm{F}^{(m)}_k} \widetilde{\bm{\partial}^{(m)}_k} [\bm{U}^{(m)}_k]^\dagger
} \,.
\end{equation}

We have therefore obtained a representation of the exact noncollinear
functional in terms of a locally collinear parameterization [\prettyref{eq:
variable_transform}]. Moreover, this relation defines the local variable
transformations required to apply locally collinear functionals to noncollinear
systems. 

\subsection{Approximate exchange--correlation functionals}

Common generalized-gradient-approximation (GGA) functionals, such as PBE
\cite{perdew:1996}, possess a comparatively simple exchange-correlation-structure in the
$u(2)^{\otimes3}$ representation space:
\begin{equation}
\begin{aligned}
\bm{F}_\mathrm{X,approx}[\bm{\rho},\bm{\gamma}] &= \bm{1}F_{\mathrm{X},0}(\widetilde{\bm{\rho}}) \mathrm{Tr}\brackets{\widetilde{\bm{\rho}}\otimes\widetilde{\bm{\gamma}}} \\\nonumber
&+ \pauli^3 F_{\mathrm{X},\mathrm{s}}(\widetilde{\bm{\rho}}) \mathrm{Tr}\brackets{\pauli_3\otimes\pauli_3\otimes\pauli_3\widetilde{\bm{\rho}}\otimes\widetilde{\bm{\gamma}}}\,, \\ 
\bm{F}_\mathrm{C,approx}[\bm{\rho},\bm{\gamma}] &= \bm{1} F_{\mathrm{C},0}(\widetilde{\bm{\rho}}) \mathrm{Tr}\brackets{\widetilde{\bm{\rho}}\otimes\widetilde{\bm{\gamma}}}\,. \\ 
\end{aligned} 
\label{eq: common_gga_fun}
\end{equation}
Correlation functionals of the Colle--Salvetti type \cite{lee:1988} may exhibit a more
intricate dependence on magnetization gradients.

Meta-GGA (mGGA) XC functionals additionally depend on the variables
$\bm{\beta}_i = \Delta\bm{\rho},\,\bm{\tau}$, where the kinetic-energy density
is given by $\bm{\tau}=\frac{\pauli_\mu}{4}\int\dots\int\mathrm{d}^3
r_2\dots\mathrm{d}^3r_N (\partial_k\Psi)^\dagger\pauli^\mu(\partial^k\Psi)$.
More generally, functionals may also be formulated in explicitly nonlocal form,
containing terms of the form 
\begin{equation}
\mathrm{Tr}\brackets{
\bm{F}^{(m+n)}_{k+l}(\bm{\rho}(\pos_1),\bm{\rho}(\pos_2)) \bm{\partial}^{(m)}_k(\pos_1) \otimes  \bm{\partial}^{(n)}_l(\pos_2)
}\,.
\end{equation}

Functionals containing traces over multiple $u(2)^{\otimes n}$ variables
require a consistent notion of spin orientation in the corresponding
eigenbasis. Consequently, eigenvalues must be ordered consistently with the
gauge convention of the underlying collinear functional. One possible
construction consists of partitioning the Bloch sphere into two hemispheres and
restricting the rotations to a single hemisphere.

\subsubsection{Relative Bloch-sphere orientation and collinear limit}
In Ref.\,\cite{pu:2023} it was pointed out that locally collinear approaches
cannot generally retain the correct collinear limit for nonlocal functionals,
because information about relative spin orientation is lost locally when using
eigenvalue densities. This statement is only correct if the relative phase of
eigenvectors is not explicitly taken into account. The relative orientation of
all eigenvectors on the Bloch sphere must therefore be taken into account
before evaluating a local, semilocal or nonlocal functional for a given point
or set of points. The correct collinear limit is retained in a globally
rotationally invariant manner, when the relative spin-frame convention of
eigenvectors $\bm{U}^{(m)}_l$ for a functional of a set of variables
$\{\bm{\beta}_l\}$ are projected on an invariant local reference vector
$\vec{d}$.

For each eigenvector $i$ the orientation on the Bloch sphere can be computed as
\begin{equation}
v^{k}_i=\brackets{[\bm{U}^{(m)}_l]^\dagger\parantheses{ \Sum{a=1}{m} 
\left(
\bigotimes_{i=1}^{a-1} \sigma^0
\right)
\otimes \sigma^k \otimes
\left(
\bigotimes_{i=a+1}^{m} \sigma^0
\right)}\bm{U}^{(m)}_l}_{ii}\,.
\end{equation}

One of the vectors $\vec{v}_i$ may then be chosen as a reference vector $\vec{d}$. The eigenvalues of all
remaining variables are then ordered according to their relative
orientation on the Bloch sphere by computing their projection on $\vec{d}$:

\begin{equation}
p_i = v_{i}^{k} d_k
\label{eq: spin_proj_dir}
\end{equation}

The eigenvectors and corresponding eigenvalues are sorted to decreasing values
of $p_i$. For a $u(2)$ variable $p_i>0\rightarrow\alpha$ and
$p_i<0\rightarrow\beta$, for a $u(2)^{\otimes2}$ variable
$\mathrm{max}(\vec{p})\rightarrow\alpha \alpha$,
$\mathrm{min}(\vec{p})\rightarrow\beta\beta$, with analogous assignments for
higher-order tensor-product spaces. 

\subsection{Approximate GGA transformations: canonical and Scalmani--Frisch variables}
\subsubsection{Canonical approach} 
The canonical approach assumes that $\tilde{\bm{\rho}}$ varies
smoothly in the local eigenbasis. The collinear gradient tensor $\widetilde{\bm{\gamma}}$ is constructed as
\begin{equation}
\widetilde{\bm{\gamma}}_\mathrm{can} = \mathrm{diag}\parantheses{\partial_k\tilde{\bm{\rho}}\otimes\partial^k\tilde{\bm{\rho}}}
\end{equation}
by direct application of the chain rule. By the Hellmann--Feynman theorem we find
\begin{equation}
\widetilde{\bm{\gamma}}_\mathrm{can} = \mathrm{diag}\parantheses{\brackets{\bm{U}_0^{\otimes 2}}^\dagger \bm{\gamma} \bm{U}_0^{\otimes2}} 
\end{equation}
$\bm{U}_0$ does not simultaneously diagonalize $\bm{\rho}$, $\bm{\gamma}$ and
the mGGA variables.  Because off-diagonal elements are neglected by
construction, the transformation cannot generally be represented as an
$SU(2)$-rotation of $\bm{\gamma}$ and the coefficient tensors appearing in the noncollinear
gradient expansion. As a result, the canonical variable replacement does not
preserve the formally energy-invariant structure of the exact transformation.
Consequently, the transformed the approach leads to a vanishing local magnetic
torque \cite{scalmani:2012}. In particular, the evaluation of functional
derivatives within canonical approach becomes ill-defined because eigenvector
derivatives are treated within an approximate projected eigenbasis rather than
the full matrix representation space. This leads to the well-known
\cite{peralta:2007,desmarais:2021} appearance of ill-defined derivatives and
unphysical properties.

\subsubsection{Scalmani--Frisch variables}
The exact tensor variable $\tilde{\bm{\gamma}}$ can be related to the collinear
variables introduced by Scalmani and Frisch \cite{scalmani:2012} by introducing
a zeroth-order tensor 
\begin{equation}
\bm{\gamma}^{(0)}=\gamma_{\mu0}
\parantheses{\mathfrak{G}^\mu\otimes\mathfrak{G}^0+(1-\delta_{\mu0})\mathfrak{G}^0\otimes\mathfrak{G}^\mu}\,.
\end{equation}
All terms bilinear in magnetization-density gradients are treated as small
perturbations. The diagonalization
$\widetilde{\bm{\gamma}^{(0)}}=[\bm{U}_{\gamma}^{(0)}]^\dagger\bm{\gamma}^{(0)}\bm{U}_{\gamma}^{(0)}$
yields four eigenvalue variables: 
\begin{align} 
\widetilde{\gamma^{(0)}}_{\pm} &= \frac{1}{4}\parantheses{\gamma_{00} \pm 2\sqrt{\gamma_{0m}\gamma^{0m}}}\,,\label{eq: sfvariables0}\\
\widetilde{\gamma^{(0)}}_{0} &= \frac{1}{4}\gamma_{00}\,,
\end{align} 

where $\widetilde{\gamma^{(0)}}_{0}$ is doubly degenerate.  The SF variables
are recovered by introducing an isotropic frozen-eigenvector
perturbation,

\begin{equation}
\bm{\gamma}'\approx\gamma_{m}{}^{m}
\bm{U}_{\gamma,0}\mathfrak{G}^3\otimes\mathfrak{G}^3\bm{U}_{\gamma,0}^\dagger\,.
\label{eq: SFPT}
\end{equation}
This is justified in the limit of near-collinearity, where all spin off-diagonal terms are small. 

The corresponding first-order corrected eigenvalues,

\begin{align}
\widetilde{\gamma^{(0)}}_{\pm}^{(1)}&= \widetilde{\gamma^{(0)}}_{\pm}+
\gamma_{m}{}^{m}/4 
\label{eq: sfvariables1}\\
\widetilde{\gamma^{(0)}}_0^{(1)}&=\widetilde{\gamma^{(0)}}_{0} -
\gamma_{m}{}^{m}/4\,,
\label{eq: sfvariables2}
\end{align}

coincide exactly with the SF variables. Projecting
$\widetilde{\gamma^{(0)}}_{\pm}$ onto $\vec{m}/\abs{\vec{m}}$---following
eq.\,\prettyref{eq: spin_proj_dir}---is equivalent to the introduction of the sign
function $f_{\nabla}$ of Ref.\,\cite{scalmani:2012}.

In contrast to $\widetilde{\bm{\gamma}}$, the mGGA variables introduced within
the SF approach coincide with those obtained in the present formalism.

\subsection{Transverse and longitudinal magnetization gradients}
Eich \emph{et al.} demonstrated that the SF approach is overly restrictive because
it treats longitudinal and transverse magnetization gradients
on an equal footing \cite{eich:2013a}. In Ref.~\cite{eich:2013a} the general noncollinear
$\bm{\gamma}$-dependent energy functional was parametrized to linear order as 

\begin{equation}
E_\mathrm{GGA}\sim\int \mathrm{d}^3r\, \alpha_0 \gamma_0 + \alpha_\parallel \gamma_\parallel + \alpha_\times \gamma_\times + \alpha_\perp \gamma_\perp\,.
\label{eq: gga_long_trans_cross}
\end{equation}

Here, the longitudinal and transverse contributions of the squared magnetization gradient are defined as

\begin{align}
\gamma_\parallel &= \hat{m}_k\hat{m}_l\gamma^{kl}\\
\gamma_\perp &= \gamma_k{}^k -  \hat{m}_k\hat{m}_l\gamma^{kl}\, 
\end{align}

where $\hat{m}=\vec{m}/\abs{\vec{m}}$ denotes the normalized magnetization
direction. In the SF approach the eigenvalues $\widetilde{\bm{\gamma_0}}$ are
perturbed by $\pm\gamma^k{}_k/4=\pm(\gamma_\parallel+\gamma_\perp)/4$ (see eq.
\prettyref{eq: sfvariables1}, \prettyref{eq: sfvariables2}), demonstrating the
absence of any distinction between longitudinal and transverse components. 

In the present formalism, the dependence of eigenvalues
$\widetilde{\bm{\gamma}}$ on $\gamma_\parallel$ and $\gamma_\perp$ is generally
nonlinear and does not admit a simple closed-form expression. To recover the
linearized form \prettyref{eq: gga_long_trans_cross}, we use first-order
perturbation theory, defining $\bm{\gamma}=\bm{\gamma}^{(0)}+\bm{\gamma}'$ and
choosing the same zeroth-order tensor $\bm{\gamma}^{(0)}$ as in the SF approach
(eq.\,\prettyref{eq: sfvariables0}).  In contrast to the SF construction,the
perturbation itself is not approximated:
\begin{equation}
\bm{\gamma}'=\gamma_{kl}\mathfrak{G}^k\otimes\mathfrak{G}^l
\end{equation}

For direct comparison with Ref.~\cite{eich:2013a}, we restrict the discussion
to exchange functionals that depend only on the perturbed eigenvalues
$\widetilde{\gamma^{(0)}}_{\pm}^{(1)}$ \prettyref{eq: common_gga_fun}.
First-order perturbation theory then yields the shifted eigenvalues:

\begin{align}
\widetilde{\gamma^{(0)}}_{\pm}^{(1)}&=\widetilde{\gamma^{(0)}}_{\pm} +
\hat{g}^k\hat{g}^l\gamma_{kl}/4\,, \nonumber\\ 
&=\widetilde{\gamma^{(0)}}_{\pm}+\frac{1}{4}\parantheses{\xi\gamma_\parallel+(1-\xi)\gamma_\perp}\,,
\label{eq: PT1}
\end{align}
where we introduced the rotationally invariant quantity
$\xi=(\hat{g}\cdot\hat{m})^2$ and
$\hat{g}_k=\gamma_{0k}/\sqrt{\gamma_{0k}\gamma^{0k}}$. $\xi$ measures the
alignment between the local magnetization direction and the spin structure of
its gradient and has values in the range $[0,1]$, where $\xi=1$ in the strictly
collinear limit.

The coefficients $\alpha_\perp$ and $\alpha_\parallel$ of
\prettyref{eq: gga_long_trans_cross} can be identified as

\begin{align}
\alpha_\parallel &= \frac{\xi}{4} &\parantheses{\frac{\delta F_\mathrm{X}[\tilde{\rho}_+,\tilde{\gamma}_+]}{\delta\gamma_+}+\frac{\delta F_\mathrm{X}[\tilde{\rho}_-,\tilde{\gamma}_-]}{\delta\gamma_-}} \\
\alpha_\perp &= \frac{(1-\xi)}{4} &\parantheses{\frac{\delta F_\mathrm{X}[\tilde{\rho}_+,\tilde{\gamma}_+]}{\delta\gamma_+}+\frac{\delta F_\mathrm{X}[\tilde{\rho}_-,\tilde{\gamma}_-]}{\delta\gamma_-}} \,.
\end{align} 

The resulting anisotropy ratio is therefore
$\alpha_\perp/\alpha_\parallel=(1-\xi)/\xi$, which correctly vanishes in the
collinear limit $\xi\rightarrow1$. 

\begin{figure}
\includegraphics[width=.5\textwidth]{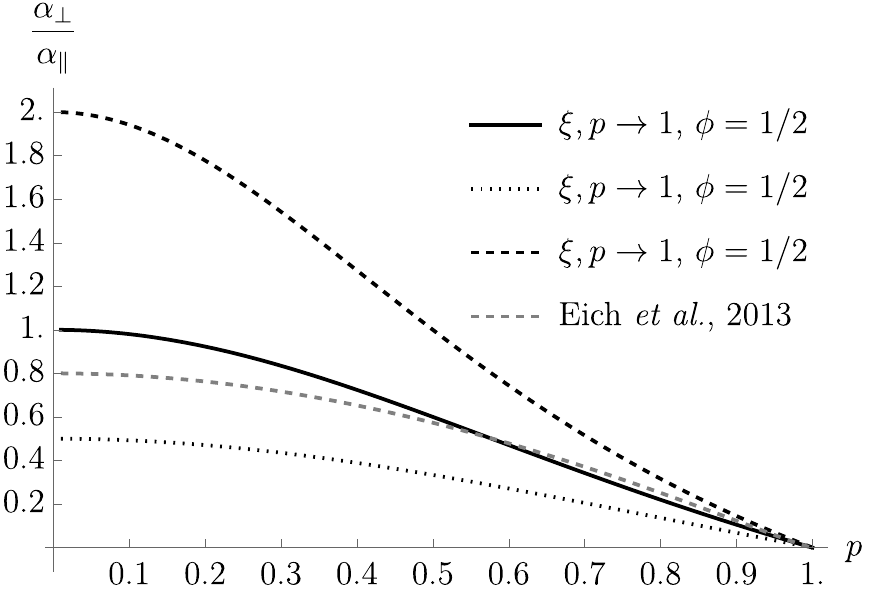}
\caption{Ratio $\alpha_\perp/\alpha_\parallel$ of linear coefficients for
transverse and longitudinal magnetization gradients as a function of the
polarization. Comparison of the approximation $\xi\sim 1-\Braket{\phi}(1-p^2)$
for representative values of $\Braket{\phi}$ with the result of
Ref.\,\cite{eich:2013a}.}
\label{fig: longitudinal_vs_transverse}
\end{figure}

For comparison with Eich \emph{et al.} \cite{eich:2013a} we can consider a
small deviation from collinearity, i.e. $\xi\approx1-\delta$, where
$0<\delta\ll1$. In the regime of large polarisation $p=\abs{\vec{m}}/\rho\sim1$
we have $\xi^{1/2}\simeq p \Rightarrow \delta\sim1-p^2$. Motivated by this
relation, we approximate $\xi\sim 1-\Braket{\phi}(1-p^2)$, where
$\Braket{\phi}$ parametrizes the effective distribution of small angular
misalignments between $\hat{m}$ and $\hat{g}$. Assuming an isotropic
distribution of relative orientations without preferred direction, a midpoint
estimate gives $\Braket{\phi}\approx1/2$, corresponding to full magnetic
isotropy at $\alpha_\perp/\alpha_\parallel\rightarrow1$. In this approximation
we obtain $\alpha_\perp/\alpha_\parallel\approx(1-p^2)/(1+p^2)$, which has the
correct collinear limit $\alpha_\perp/\alpha_\parallel\rightarrow0$. In
\prettyref{fig: longitudinal_vs_transverse} we compare
$\alpha_\perp/\alpha_\parallel$ for angular distributions
$\Braket{\phi}=1/2,\,1/3,\,2/3$ with the result of Ref.\,\cite{eich:2013a}. The
present estimate for $\Braket{\phi}=1/2$ agrees to within approximately 20\,\%
with the polarisation dependence derived in Ref.\,\cite{eich:2013a} for
$p>0.9$.      

We emphasize that, within the present formalism, the nontrivial dependence on
longitudinal and transverse magnetization gradients emerges naturally from the
geometrical structure of the magnetization density itself and does not need to
be introduced explicitly at the level of the XC functional.

\subsection{Noncollinear functional derivatives: potentials, kernels and higher order derivatives}

Numerically robust functional derivatives are essential for practical
applications, for example within a Kohn--Sham (KS) approach. Whereas
self-consistent field (SCF) KS requires only first derivatives of the
XC-functional (XC-potential) are required, linear response theory---required
for time-dependent DFT and second order properties---as well as calculations of
local magnetic torques require second order derivatives (XC-kernel).
Accordingly, higher-order response theory requires the $n$-th functional derivatives of
the potential ($n$th-order hyperkernels).

\subsubsection{Transformation of collinear functional derivatives to noncollinear space}
A practical advantage of the present formalism is that noncollinear derivatives are
obtained directly through unitary transformation of the eigenspace derivatives.
Once the full $n$th-order derivative tensor in eigenspace $\bm{D}^{n}_{i_1\dots
i_n}$ is formulated, the corresponding noncollinear derivatives follow immediately:

\begin{equation}
\frac{\delta^n F}{\delta\bm{\beta}_{i_1}\dots\delta\bm{\beta}_{i_n}} =
\bm{U}^{(\sum_k d_k)}_{i_1\dots i_n} 
\bm{D}^{(n)}_{i_1\dots i_n}
\brackets{\bm{U}^{(\sum_k d_k)}_{i_1\dots i_n}}^\dagger\,.
\label{eq: functional_derivs}
\end{equation}

The $n$-th order eigenspace derivative matrix $\bm{D}^{(n)}_{i_1\dots i_n}$ has
dimension $2^{\sum_k d_k}$, where $d_k$ denotes the tensor order associated
with the representation space. For practical applications, the noncollinear
derivative can be expanded in the generator basis of the joint variable space
$\bigotimes_k u(2)^{\otimes d_k}$ as
\begin{equation}
\frac{\delta^n F}{\delta\beta_{i_{1},\vec{\mu}_{i_1}}\dots\delta\beta_{i_{n},\vec{\mu}_{i_n}}} 
=\mathrm{Tr}\brackets{
\bm{O}^{(\sum_k d_k)}_{\vec{\mu}_{i_1}\dots\vec{\mu}_{i_n}}
\bm{D}^{(n)}_{i_1\dots i_n}
}
\,.
\label{eq: functional_derivs_pauli}
\end{equation}

We define the $\bigotimes_k u(2)^{\otimes d_k}$ noncollinear density operator

\begin{equation}
\bm{O}^{(\sum_k d_k)}_{\vec{\mu}_{i_1}\dots\vec{\mu}_{i_n}}
=
\brackets{\bm{U}^{(\sum_k d_k)}_{i_1\dots i_n}}^\dagger
\widetilde{\bm{O}}^{(\sum_k d_k)}_{\vec{\mu}_{i_1}\dots\vec{\mu}_{i_n}}
\bm{U}^{(\sum_k d_k)}_{i_1\dots i_n}\,.
\end{equation}

and the corresponding operator in eigenspace
\begin{equation}
\widetilde{\bm{O}}^{(\sum_k d_k)}_{\vec{\mu}_{i_1}\dots\vec{\mu}_{i_n}}
=
\brackets{\bigotimes_{k=1}^n\parantheses{\frac{1}{d_k!}\Sum{p_k\in S_{d_k}}{}\bigotimes_{\nu\in p_k(\vec{\mu}_{i_k})}\bm{\mathfrak{G}}^{\nu}}}^\dagger\,,
\end{equation}
where $p_k\in S_{d_k}$ denotes an element of the permutation group of degree $d_k$.

\subsubsection{Eigenspace derivatives}

The first-order eigenspace derivative tensor $\bm{D}^{(1)}$ is diagonal and can be constructed from the collinear functional
derivatives with respect to eigenspace variable $\tilde{\bm{\beta}}$ in
$u(2)^{\otimes m}$ representation space as 
\begin{equation}
(\bm{D}^{(1)})_{\vec{a}\vec{b}} = 
\delta_{\vec{a}\vec{b}}\frac{\delta
\tilde{F}}{\delta\tilde{\beta}_{\vec{a}}}\,,
\end{equation}
where $\delta_{\vec{a}\vec{b}}=\prod\limits_{k=1}^m\delta_{a_k}\delta_{b_k}$
and indices $\vec{a},\vec{b},\vec{c},\vec{d}$ represent spin configurations of 
$m$ spin-$1/2$ degrees of freedom.

Using the spectral representation of Fr\'echet derivatives for matrix
functions, the corresponding eigenspace derivatives can be expressed in terms
of divided differences. In practical terms, the first-order divided difference
matrix of the functional in eigenspace with respect to $\tilde{\bm{\beta}}$ is

\begin{equation}
\mathcal{D}_{\vec{a}\vec{b}}^{(1)}(\tilde{\bm{\beta}}) =
\begin{cases}
\frac{\frac{\delta \tilde{F}}{\delta
\tilde{\beta}_{\vec{a}}}-\frac{\delta
\tilde{F}}{\delta\tilde{\beta}_{\vec{b}}}}{\tilde{\beta}_{\vec{a}}-\tilde{\beta}_{\vec{b}}}
& \tilde{\beta}_{\vec{a}} \not= \tilde{\beta}_{\vec{b}}\\
\frac{1}{2}\parantheses{\frac{\delta^2
\tilde{F}}{\delta\tilde{\beta}_{\vec{a}}^2}+\frac{\delta^2
\tilde{F}}{\delta\tilde{\beta}_{\vec{b}}^2}} - \frac{\delta^2
\tilde{F}}{\delta\tilde{\beta}_{\vec{a}}\delta\tilde{\beta}_{\vec{b}}}  &
\tilde{\beta}_{\vec{a}} = \tilde{\beta}_{\vec{b}}\\
\end{cases}\,.
\end{equation}
The divided-difference functions remain smooth in the limit
$\tilde{\beta}_{\vec{a}}-\tilde{\beta}_{\vec{b}}=0$. In numerical calculations
this smoothness can be retained with a smooth switching function as described
in appendix \ref{sec: numdivdif}.

The eigenspace Hessian matrix with respect to variable
$\bm{\beta}$ is:
\begin{equation}
\begin{aligned}
(\bm{D}^{(2)})_{\vec{a}\vec{b},\vec{c}\vec{d}} &= 
\delta_{\vec{a}\vec{b}}\delta_{\vec{c}\vec{d}}\frac{\delta^2
\tilde{F}}{\delta\tilde{\beta}_{\vec{a}}\delta\tilde{\beta}_{\vec{c}}} 
+ (1-\delta_{\vec{a}\vec{b}})\delta_{\vec{a}\vec{d}}\delta_{\vec{b}\vec{c}}
\mathcal{D}_{\vec{a}\vec{b}}^{(1)}\,.
\end{aligned}
\end{equation}

For a GGA functional, the second order derivatives with respect to
$\tilde{\bm{\rho}}$ and $\tilde{\bm{\gamma}}$ are 
\begin{align}
(\bm{D}^{(2)}(\tilde{\bm{\rho}}))_{ab,cd} &= 
\delta_{ab}\delta_{cd}\frac{\delta^2 \tilde{F}}{\delta\tilde{\rho}_{a}\delta\tilde{\rho}_{c}} 
+ (1-\delta_{ab})\delta_{ad}\delta_{bc}\mathcal{D}_{ab}^{(1)}(\tilde{\bm{\rho}})\,,
\end{align}
and
\begin{align}
&(\bm{D}^{(2)}(\tilde{\bm{\gamma}}))_{a_1a_2b_1b_2,c_1c_2d_1d_2} \\\nonumber
&= 
\delta_{a_1b_1}\delta_{a_2b_2}\delta_{c_1d_1}\delta_{c_2d_2}\frac{\delta^2
\tilde{F}}{\delta\tilde{\gamma}_{a_1a_2}\delta\tilde{\gamma}_{c_1c_2}} \\&+
(1-\delta_{a_1b_1}\delta_{a_2b_2})\delta_{a_1d_1}\delta_{a_2d_2}\delta_{b_1c_1}\delta_{b_2c_2}
\mathcal{D}_{a_1a_2b_1b_2}^{(1)}\,.
\end{align}

The tensor structure of mixed second-order derivatives follows directly from
tensor products of the corresponding first order derivative tensor.

Existing collinear XC functionals are formulated under the symmetry constraint
$\widetilde{\gamma}_{\alpha\beta}=\widetilde{\gamma}_{\beta\alpha}$. In full
$u(2)^{\otimes2}$ representation this is, however, not generally true.
Therefore, eigenspace variables and derivative tensors have to be transformed
into a symmetrized space representation. This achieved by introducing the
unitary transformation 

\begin{equation}
\bm{U}_\mathrm{sym} = \frac{1}{\sqrt{2}}\begin{pmatrix} 
1 &0 &0 &0\\
0 &1 &1 &0\\
0 &1 &-1 &0\\
0 &0 &0 &1\\
\end{pmatrix}\,,
\end{equation}
which rotates
$\overline{\bm{\gamma}}=\bm{U}_\mathrm{sym}^\dagger\widetilde{\bm{\gamma}}\bm{U}_\mathrm{sym}$
and $\overline{\bm{U}_\gamma}=\bm{U}_\mathrm{sym}\bm{U}_\gamma$ to the basis of
available collinear functionals. This transformation constitutes an
approximation because it restricts the form of representable noncollinear
functionals.  It remains exact, however, for exchange functionals, which are
not functionals of $\gamma_{\alpha\beta}$ and GGA functionals of form
\prettyref{eq: common_gga_fun} like PBE. 

Higher-order eigenspace derivative tensors can be constructed recursively from
the lower-order derivative tensors through recursive divided-difference
constructions.

\section{Numerical results}
\subsection{Implementation and computational details}
We implemented the present $SU(2)$ rotation-based approach to noncollinear DFT
within the quasi-relativistic two-component DFT code
\cite{wullen:1998,wullen:2002,gaul:2020a,colombojofre:2022,bruck:2023} of the
program package described in Ref.~\cite{wullen:2010}, which is based on the
integral engine of Turbomole \cite{ahlrichs:1989}. For comparison, we also
implemented the XC potential and kernel for the canonical approach. For
internal testing, we additionally implemented a chain-rule based version of the
SF approach. 

The local magnetic torque $\vec{t}$ was evaluated at every point $\pos$ as:
\begin{equation}
\begin{aligned}
t_j &= \epsilon_{ijk} \rho^i \brackets{\frac{\delta
F_\mathrm{XC}}{\delta\rho_j} - \partial_k \frac{\delta
F_\mathrm{XC}}{\delta\partial_k\rho_j}} \\
&= \epsilon_{ijk} \rho_i \brackets{\frac{\delta F_\mathrm{XC}}{\delta\rho_j} 
- 2\frac{\delta F_\mathrm{XC}}{\delta\gamma_{j\mu}} (\partial_k\partial^k\rho_\mu) 
\right.\\ &\left. 
- 2\frac{\delta^2 F_\mathrm{XC}}{\delta\rho_\nu\delta\gamma_{j\mu}} (\partial_k\rho_\nu)(\partial^k\rho_\mu) 
\right.\\ &\left. 
- 4\frac{\delta^2 F_\mathrm{XC}}{\delta\gamma_{\nu\kappa}\delta\gamma_{j\mu}} (\partial^l\rho_{\nu})(\partial_l\partial_k\rho_{\kappa})(\partial^k\rho_\mu)}\,,
\end{aligned}
\end{equation}
where we exploited the symmetry of $\gamma_{\mu\nu}$.

\subsection{Local magnetic torque in \ce{Cr3}}
To facilitate comparison with previous studies of noncollinear extensions of DFT, we
used the molecular structure of the \ce{Cr3} cluster reported in
Refs.\,\cite{peralta:2007,scalmani:2012,pu:2023}, namely a bond lengths of
$3.7\,a_0$ and $D_{3h}$ symmetry. We oriented the molecule in the $x$--$z$ plane
and constrained the magnetization density to lie within this plane. We note that
the generalized Hartree--Fock (GHF) or Kohn--Sham (GKS) solutions correspond to
real solutions as discussed in Ref.\,\cite{stuber:2003}. Rotating to the
$x$--$y$ or $y$--$z$ plane leads to complex solutions. All three correspond to different 
realizations of the magnetic group introduced by Fukutome \cite{fukutome:1981}.
We used a valence basis set together with a small-core scalar relativistic effective
core potential by Dolg \emph{et al.} \cite{dolg:1987}, which was employed in
previous studies \cite{peralta:2007,scalmani:2012,pu:2023}. 

To obtain an initial guess converging to the broken spin-symmetry solution of
the GHF and GKS equations we applied local pseudo-magnetic fields with a
strength of $10 \mathrm{a.u.}$ within a sphere of radius $1\,a_0$ around each
Cr atom, which pointed in three different directions and coupling only to the
spin angular momentum. This was realized by adding a Hamiltonian of form
$\pauli^3\Theta(\pos-\pos_1)+(\pauli^1+0.1\pauli^3)\Theta(\pos-\pos_2)+(-\pauli^1+0.1\pauli^3)\Theta(\pos-\pos_3)$.
The field strength was gradually reduced until becoming zero. Starting from the
converged cGHF wave function, the cGKS equations were solved self-consistently
for various functionals, including LDA and hybrid LDA
\cite{dirac:1930,slater:1951,vosko:1980,becke:1993}, as well as various GGA
functionals (PBE \cite{perdew:1996}, BLYP \cite{becke:1988,lee:1988}, OLYP
\cite{handy:2001,lee:1988}, KT3 \cite{keal:2004}, PBE0 \cite{adamo:1999}). All
calculations were converged until the relative energy change between successive
SCF iterations was below $10^{-9}\,E_\mathrm{h}$.

Following previous studies\,\cite{peralta:2007,scalmani:2012,pu:2023}, the
electronic configurations are characterized by atomic magnetic moments
$m=\abs{\Braket{\vec{\pauli}}_\mathrm{atom}}\mu_\mathrm{B}$ obtained from a
Mulliken population analysis of the magnetization density and the expectation
value of the squared spin operator $\Braket{\hat{S}^2}$. The latter was
computed for the GHF and GKS determinants, including contributions from
nonlocal exchange
\begin{equation}
\Braket{\hat{S}^2} =\frac{3\hbar^2}{4} N + \mathrm{Tr}\brackets{\bm{S}_{k}}\mathrm{Tr}\brackets{\bm{S}^{k}} - \mathrm{Tr}\brackets{\bm{S}_{k}\bm{S}^{k}}\,,
\end{equation}
where $N$ is the number of electrons, $\vec{\bm{S}}$ denotes the spin-density matrix in single-particle space.

In the present article we restrict our discussion to LDA, GGA and hybrid
functionals. We leave applications to mGGA and nonlocal functionals for future
work.

\begin{figure*}
\includegraphics[width=\textwidth]{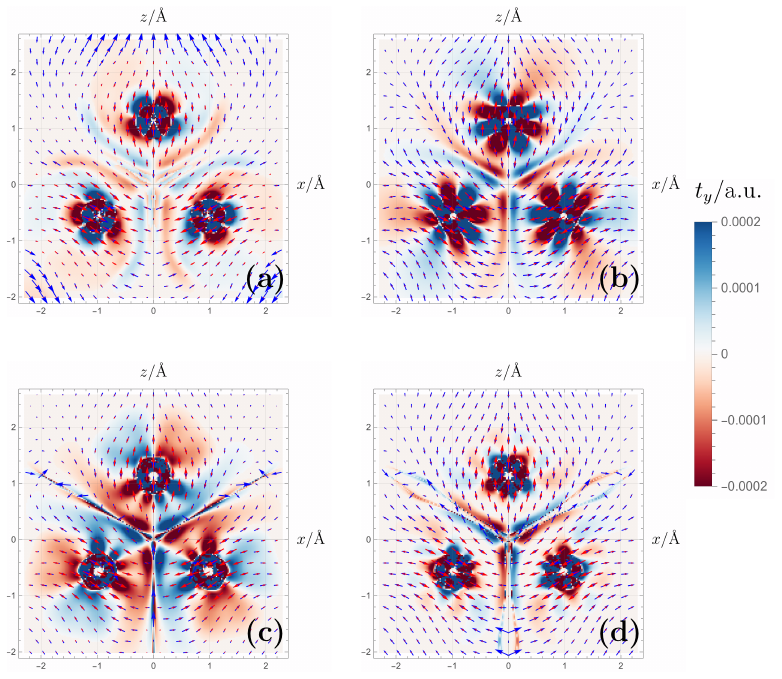}
\caption{Magnetization (red), XC magnetic field (blue) and local magnetic
torque (heat map) in \ce{Cr3} oriented in the $x$--$z$ plane. The magnetization
density was oriented in the same plane. All calculations were performed with
the PBE XC-functional. We show different levels of approximation to
$\bm{\gamma}$: the lowest order solution $\widetilde{\bm{\gamma}^{(0)}}$ based
on the Kohn-Sham orbitals of NC-$SU(2)$ (a), the NC-$SU(2)$-SF
[eq.\,\prettyref{eq: SFPT}] (b), the NC-$SU(2)$-PT1
approximation [eq.\,\prettyref{eq: PT1}] (c) and the full NC-$SU(2)$ (d).}
\label{fig: torque1}
\end{figure*}

\begin{figure*}
\includegraphics[width=\textwidth]{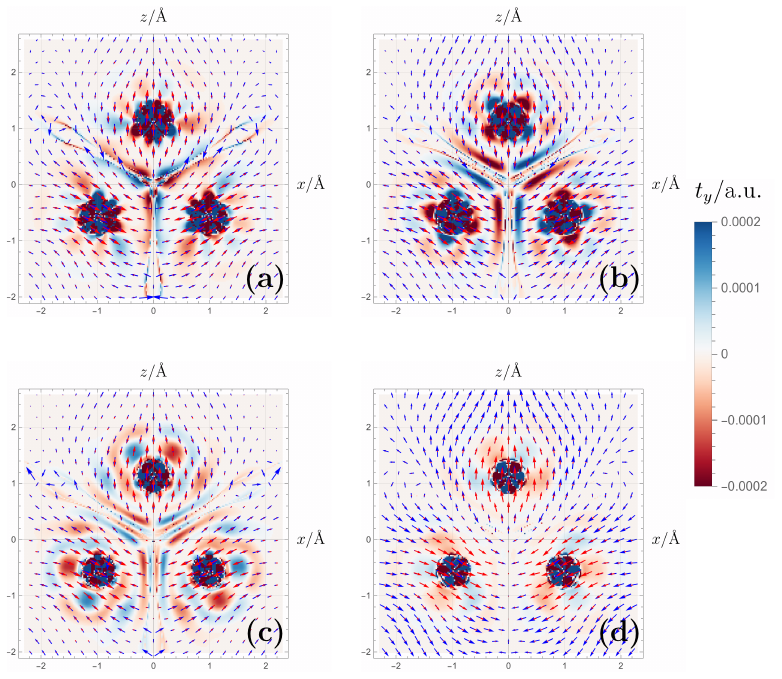}
\caption{Magnetization (red), XC magnetic field (blue) and local magnetic
torque (heat map) in \ce{Cr3} oriented in the $x$--$z$ plane. The magnetization
density was oriented in the same plane. We show results with the NC-$SU(2)$
approach for different XC functionals: BLYP (a), OLYP (b), KT3 (c) and the
correlation torque from LYP (d).}
\label{fig: torque2}
\end{figure*}

All results for $m$ and $\Braket{\hat{S}^2}$ are summarized in Table
\ref{tab:cr3}. In Figures \ref{fig: torque1} and \ref{fig: torque2} we show the
magnetization density, the effective XC magnetic field and the local
magnetic torque perpendicular to the plotting plane. 

\begin{table*}
\caption{Squared spin operator expectation value and atomic magnetic moments in
\ce{Cr3} obtained from different approaches to noncollinear DFT employing
different GGA XC-functionals and an ECP basis by Dolg \cite{dolg:1987}. The results are
compared to previous calculations with the same molecular structure and basis
set. Result obtained with the approach derived in this work are denoted NC-$SU(2)$,
NC-$SU(2)$-PT1 for the first-order perturbation theory in $\bm{\gamma}'$
[eq.\,\prettyref{eq: PT1}] and NC-$SU(2)$-SF using an approximate first-order
perturbation [eq. \prettyref{eq: SFPT}] to reproduce the SF
approach \cite{scalmani:2012}.} 
\label{tab:cr3}
\begin{tabular}{cccccccccccccccc}
\toprule
&\multicolumn{7}{c}{$\Braket{\hat{S}^2}/\hbar^2$} && \multicolumn{7}{c}{$m_{\ce{Cr}}/\mu_\mathrm{B}$} \\
\cline{2-8} \cline{9-16}
&LDA & PBE & BLYP & OLYP & KT3 & PBE0 & GHF &&LDA & PBE & BLYP & OLYP & KT3 & PBE0 & GHF \\
\midrule
Canonical                    &\multirow{4}{*}{3.27}&3.88&3.40&4.44&5.12&6.33&\multirow{4}{*}{8.11} & &\multirow{4}{*}{2.89}&3.32&2.91&3.61&4.06&4.80&\multirow{4}{*}{5.91}\\
NC-$SU(2)$-SF                &                     &3.90&3.43&4.44&5.09&6.36& &                      &                     &3.42&3.02&3.72&4.11&4.88&\\
NC-$SU(2)$-PT1               &                     &3.91&3.39&4.46&5.15&6.37& &                      &                     &3.38&2.95&3.65&4.09&4.84&\\
NC-$SU(2)$                   &                     &3.93&3.42&4.46&5.15&6.35& &                      &                     &3.38&2.96&3.63&4.07&4.80&\\
\midrule
Canonical \cite{peralta:2007}&3.25&3.87&&&&6.32&8.11& &2.88 & 3.32 &&&& 4.80 & 5.90\\
SF\cite{scalmani:2012} & - & 3.89 \\
Multi-collinear\cite{pu:2023}&3.25&3.89&&&&6.36&8.11& &2.88 & 3.43 &&&& 4.89 & 5.90\\
\bottomrule
\end{tabular}
\end{table*}

We can successfully reproduce the cGHF and cGKS-LDA values of
Refs.\,\cite{peralta:2007,pu:2023} for the canonical approach as well as the
PBE values for the SF approach reported in Ref.\,\cite{scalmani:2012}. We
observe small variations in $\braket{\hat{S}^2}$ and $m$ across different
noncollinear approximations. The NC-$SU(2)$ approach yields slightly larger
values of $\Braket{\hat{S}^2}$ compared to the SF approximation for PBE, OLYP
and KT3, whereas for BLYP and PBE0 this behavior is not observed. For the
magnetic moment $m$, values obtained with the NC-$SU(2)$ approach are smaller
than those obtained within the SF approximation.  

The local torque structure with the PBE functional is shown in
\prettyref{fig: torque1} for \ce{Cr3} being oriented in the $x$--$z$ plane. 

We tested rotational invariance by computing the local torques for
different orientations of Cr$_3$. We always obtained the same result for
the local torque structure and the total energies remained invariant within numerical
precision. Moreover, the zero-torque theorem \cite{capelle:2001} is fulfilled
by the present approach. The net perpendicular torque was below
$10^{-5}\,\mathrm{a.u.}$, once the norm of the difference density had
converged to the same numerical precision. We found that in analogy to the
gradient of the charge density the integrated torque provides a helpful measure
for convergence of the magnetization density. This is expected since the torque
is the net magnetic force, which should vanish at the minimum
\cite{capelle:2001}.

We can reproduce the local torque found in Ref.\,\cite{scalmani:2012} with the
$SU(2)$-SF approach [\prettyref{fig: torque1} (b)]. A similar torque structure
was found by Pu \emph{et al.} with their multicollinear approach
\cite{pu:2023}. 

In contrast to the SF approach, the NC-$SU(2)$-PT1 [\prettyref{fig:
torque1} (c)] and NC-$SU(2)$ $\bm{\gamma}$ [\prettyref{fig: torque1} (d)] lead
to a qualitatively different torque structure with only four nodes around each Cr
atom. In the zeroth order approximation, i.e.\ when neglecting all
$\gamma_{kl}$, the same four node structure is observed [\prettyref{fig:
torque1} (a)]. The eight-fold phase change observed in the SF approach and the
multicollinear approach likely originates from the assumption of full isotropy
in the $\gamma_{kl}$-perturbation. The NC-$SU(2)$ approach leads to a
torque structure that closely resembles the torque structure found with the
spin-current dependent noncollinear mGGA approaches or exact-exchange
functionals \cite{tancogne-dejean:2022}. However, the torque structure obtained
within the present approach exhibits additional phase changes in the interstitial region
than found with an exact exchange functional being in this respect closer to
the Slater-exchange model (see Slater-KLI and EXX-KLI,
Ref.\,\cite{tancogne-dejean:2022}). We note that already at first-order, we observe
a strongly increased torque between the Cr atoms featuring a long-range tail
which differs considerably from the SF approach. 

One important limitation of PBE is its lack of magnetization-gradient dependence
in the correlation part. Only few correlation functionals are explicit
functionals of the magnetization gradient. One frequently-used functional with
this feature was developed by Lee, Yang and Parr (LYP) \cite{lee:1988}. We
compare the magnetic torque description of different functionals employing LYP
correlation in \prettyref{fig: torque2}. We choose the widely applied BLYP
functional, the optimized exchange functional by Handy and Cohen
\cite{handy:2001} (OLYP) and the semi-empirical KT3 functional
\cite{keal:2004}, which was developed for the accurate description of magnetic
properties. KT3 may therefore be expected to provide an improved description of
spin densities. We find a strong functional dependence of the torque structure.
Whereas the differences between PBE and BLYP can largely be attributed to the
additional correlation torque, the OLYP and KT3 functionals lead to torque
structures which are much closer to the EXX-KLI approximation
\cite{ullrich:2018,tancogne-dejean:2022}. In the case of the KT3 functional the
main difference to the EXX-KLI torque is a richer phase structure in the
interstitial region. This implies that the choice of the functional can have
a significant influence on real-time dynamics simulations of noncollinear
systems.

Although we find for all functionals eight nodes around the Cr atoms for the SF
approximation (see Supplementary Material), we note that for the OLYP functional
the additional phase changes exhibit smaller amplitudes. In general, we find
that the variation between different noncollinear approximations is least
pronounced for OLYP.

In summary, the effects from the exact treatment of noncollinearity on
$\braket{\hat{S}^2}$ and atomic magnetic moments are mild but are crucial
for accurately describing the local magnetic torque structure. The magnetic
torque computed with the KT3 functional shows the best agreement with dedicated
noncollinear mGGA functionals, as developed in
Ref.\,\cite{tancogne-dejean:2022}. We expect that the present approach will
pave the way to accurate real-time spin-dynamics simulations of noncollinear systems
with commonly available collinear functionals.  

\subsection{Influence of noncollinear approximations on hyperfine coupling in uranium}

We computed the electronic ground states of atomic U and
\ce{U+}. Uranium features strongly spin-orbit-coupled high spin states, with
approximate term symbols $^5\mathrm{L}_6$ and $^4\mathrm{I}_{9/2}$, respectively. We employ
the same methodology as in Ref.\,\cite{stricker:2025a}. Relativistic effects
were considered on the two-component zeroth order regular approximation
(2c-ZORA) level with a damped model potential to alleviate
the gauge dependence \cite{wullen:1998,liu:2002}.
The nuclear charge density distribution was
modeled as a normalized spherical Gaussian 
$\varrho_K \left( \vec{r} \right) = \frac{\zeta_K^{3/2}}{\pi ^{3/2}}
\text{e}^{-\zeta_K \left| \vec{r} - \vec{r}_K \right| ^2}$ with $\zeta_K =
\frac{3}{2 r^2 _{\text{nuc},K}}$. The root-mean-square radius
$r_{\text{nuc},K}$ was chosen as suggested by Visscher and Dyall employing the 
$^{238}$U isotope \cite{visscher:1997}. 
 Uranium was modeled with an atom-centered Gaussian basis set using the doubly-augmented 
core-valence basis set of triple-$\zeta$ quality by Dyall (d-aug-dyall.cv3z)
\cite{dyall:2002, dyall:2006}. Electronic densities were converged until the
change of the total energy between two consecutive cycles in the
self-consistent field procedure was below $10^{-10}\,E_\mathrm{h}$.

Kohn-Sham spinors are computed with different hybrid versions of OLYP, which
was found to be most robust to different noncollinear approximations among the
studied functionals for \ce{Cr3}. For comparison we employ GHF and hybrid LDA
with 20\,\%{}, 40\,\%{} and 50\,\%{} Fock exchange \cite{becke:1993}.  We note,
that we were not able to converge to a stable cGKS solution with pure,
non-hybrid XC functionals for the studied high angular momentum states. This
behavior could be reproduced with a GHF calculation without current densities.
Therefore, we conclude that spin-current densities must be explicitly included
in the exchange functional to converge pure cGKS calculations for these
systems. 

We computed the ionization energy and report approximate spin-quantum numbers
connected with $\Braket{\hat{S}_z}$, $\Braket{\hat{S}^2}$. Finally, employing
our toolbox approach for two-component molecular properties \cite{gaul:2020} we
compute the magnetic dipole hyperfine coupling constant of \ce{^{233}U} and \ce{^{233}U+} (see also
\cite{gaul:2020a}) as
\begin{equation}
A = \frac{\mu(^{233}\mathrm{U})\Braket{\hat{H}_\mathrm{hf}}}{J I(^{233}\mathrm{U})},
\end{equation}
where $J$ is the reduced total electronic angular momentum and the nuclear
magnetic dipole moment and total nuclear angular momentum of \ce{^{233}U} are
$\mu(^{233}\mathrm{U})=0.59(10)$ and $I(^{233}\mathrm{U})=5/2$
\cite{stone:2005}. Details on the definition of $\hat{H}_\mathrm{hf}$ can be
found in Appendix \ref{sec: hyperfine}. All properties are listed in
\prettyref{tab: uranium} for different levels of noncollinear GGA, NC-$SU(2)$,
NC-$SU(2)$-PT1, NC-$SU(2)$-SF, and the canonical approach.  The total electronic angular
momentum was aligned approximately along the $z$ axis. 

The influence of the noncollinear approximation on the ionization energy is
small. The largest difference is $\sim0.1\,\mathrm{eV}$ for the SF
approximation. The influence of the chosen noncollinear model will likely be
larger for pure semilocal XC functionals, which could not be applied to the
uranium system. From the present data, we conclude that the choice of a
noncollinear model is negligible compared to the overall influence of the GGA
corrections to LDA. 

In contrast, hyperfine coupling constants $A$ show pronounced sensitivity to
the treatment of noncollinearity. We discuss mainly the results for 40\,\%{}
Fock exchange because the influence of the XC functional decreases with
increasing Fock exchange. For neutral uranium, the difference between the full
first-order approximation and the exact diagonalization of $\bm{\gamma}$ is
negligible ($<1\,\%$). The difference between the SF approximation and the full
$\bm{\gamma}$ exceeds $20\,\%{}$. This effect is larger than
the influence of the OLYP correction to LDA ($15\,\%$). The influence of Fock
exchange is only subtle (2\,\%). The canonical result is comparatively close to
the approach with exact $SU(2)$-rotation, agreeing generally much better with
the full $SU(2)$ approach than the SF approach. By comparison with experiment
we see large deviations, indicating that the large amount of Fock exchange,
which was applied, is not appropriate for a reliable description of the
hyperfine coupling of uranium.
 
In the uranium ion, correlation effects on $A$ appear to be less important. The
difference between HF and LDA40 is less than 10\,\%. Consequently the influence
of the noncollinearity model is less pronounced. In contrast to the
observation for neutral uranium, here the canonical approach shows most
pronounced deviations from the full $SU(2)$ approach ($\sim3\,\%$), which are
of the same size as the total effect of the OLYP-GGA correction to LDA. We note
here that the canonical approach was converging much slower than the other
approaches.

We emphasize that only a single GGA functional was employed in the present
study.  The observations indicate that the choice of noncollinear model can
lead to significant differences in magnetic properties, comparable in magnitude
to differences between XC functionals themselves. It may be worthwhile to
investigate these effects in systems where pure XC functionals can be applied.
We anticipate that the influence of the noncollinearity model may be even more
pronounced for pure functionals.

\begin{table}
\caption{$\Braket{\hat{S}_z}$, $\Braket{\hat{S}^2}$, the magnetic dipole
hyperfine coupling constant $A$ of \ce{^{233}U} and \ce{^{233}U+}, and the
ionization energy $I$ of \ce{^{233}U}. All  values are computed at the level of
2c-ZORA-cGKS/d-aug-dyall.cv3z, employing different functionals and different
noncollinear approaches. $I$ and $A$ are compared to experimental (exp.) and
theoretical (theo.) data in the literature. We employed
$\mu(^{233}\mathrm{U})=0.59\,\mu_\mathrm{N}$ and $I(^{233}\mathrm{U})=5/2$
\cite{stone:2005} in calculations of $A$.}
\label{tab: uranium}
\begin{tabular}{lS[round-precision=2,round-mode=places,table-format=1.2]S[round-precision=2,round-mode=places,table-format=1.2]S[round-precision=4,round-mode=figures,table-format=-3.2(2)]S[round-precision=4,round-mode=figures,table-format=1.5(1)]}
\toprule
Method & {$S_z/\hbar$} & {$\Braket{\hat{S}^2}/\hbar^2$}& {$A$/MHz} &{$I/\mathrm{eV}$} \\
\midrule
\ce{U},$^{5}\mathrm{L}_{6}$ \\
\midrule
HF                     & 2.17304 & 8.15351 & -46.5023     & 5.979850603431048\\
\\
\midrule
\multicolumn{3}{c}{60\,\% Fock exchange} \\
\midrule
OLYP-$SU(2)$          & 1.7138  & 5.77304 &  96.8096   & 6.067048057043415\\
OLYP-$SU(2)$-PT1      & 1.69931 & 5.73683 &  96.5454   & 6.067166234243191\\
OLYP-$SU(2)$-SF       & 1.68732 & 5.86261 &  76.0193   & 6.1198045831793895\\
OLYP-Can.             & 1.71089 & 5.75856 &  96.7538   & 6.058658530135191\\
LDA                   & 1.68074 & 5.7018  &  98.0341   & 6.472360938717207\\
\midrule
\multicolumn{3}{c}{40\,\% Fock exchange} \\
\midrule
OLYP-$SU(2)$         & 1.70847 & 5.77005 &  98.4748  & 5.976442814708331\\
OLYP-$SU(2)$-PT1     & 1.68411 & 5.71002 &  98.0827  & 5.977213106389954\\
OLYP-$SU(2)$-SF      & 1.64896 & 5.82517 &  72.6465  & 6.071139429442864\\
OLYP-Can.            & 1.70387 & 5.76007 &  97.8747  & 5.972474691714109\\
LDA                  & 1.64187 & 5.54996 & 112.722   & 6.3549280288133625\\
\\
exp.\cite{coste:1982,gangrskij:1996}   &         &         &  131.56(10) & 6.19405(6) \\
\midrule
\ce{U+},$^{4}\mathrm{I}_{9/2}$\\
\midrule
HF                     & 1.33539 & 3.82545 & 157.344    & \\
\\
\midrule
\multicolumn{3}{c}{60\,\% Fock exchange} \\
\midrule
OLYP-$SU(2)$          & 1.28757 & 3.70708 &   164.77  \\
OLYP-$SU(2)$-PT1      & 1.28407 & 3.70034 &   162.822  \\
OLYP-$SU(2)$-SF       & 1.27278 & 3.67927 &   163.022  \\
OLYP-Can.             & 1.28907 & 3.7291  &   170.2  \\
LDA                   & 1.25914 & 3.65121 &   166.708  \\
\midrule
\multicolumn{3}{c}{40\,\% Fock exchange} \\
\midrule
OLYP-$SU(2)$         & 1.28594 & 3.70747 &  167.243 \\
OLYP-$SU(2)$-PT1     & 1.2797  & 3.6907  &  164.463 \\
OLYP-$SU(2)$-SF      & 1.26301 & 3.6615  &  164.925 \\
OLYP-Can.            & 1.28579 & 3.70295 &  163.136 \\
LDA                  & 1.24388 & 3.62589 &  170.283 \\
\\
theo.\cite{porsev:2022}    &          &         &  137(10)  \\
\bottomrule
\end{tabular}
\end{table}

\section{Conclusion}
We presented a rigorous approach to noncollinear DFT with semilocal collinear
functionals, which follows from the gradient expansion of a formally exact
noncollinear functional. The present approach fulfills all key requirements
identified in previous works \cite{eich:2013a,pu:2023}, can be
straightforwardly implemented, and is numerically robust. We demonstrated that
the canonical and Scalmani--Frisch approaches emerge as approximations within
the present formalism. Using the \ce{Cr3} cluster as a prototypical example, we
showed that a generalization of the GGA exchange-functionals KT3 and OPTX
within the present formalism can qualitatively reproduce local magnetic torque
structures of dedicated spin-current mGGA and exact exchange calculations.

For the hyperfine coupling constants of U and \ce{U+} we could
reveal an unprecedented dependence of magnetic properties on the treatment of
noncollinearity. These results can provide valuable insight into the requirements
for constructing XC functionals that are capable of accurately describing
magnetic properties in strongly spin-orbit-coupled systems.

We anticipate that the present approach will be particularly beneficial for
extensions to noncollinear spin-current DFT as it is general for
$SU(2)$-invariant variables, and for higher order response calculations because
functional derivatives in eigenbasis representation can be computed
recursively. The present work provides a foundation for improved descriptions
of magnetic properties in high-spin states, particularly in actinide compounds.

\begin{acknowledgments}
This work was generously funded by the Fonds der Chemischen Industrie through a
Liebig fellowship.
\end{acknowledgments}

\appendix

\section{Regularization for vanishing magnetization}
Although the present approach is well-defined in all density regions, care has
to be taken when approaching the limit of vanishing magnetization, when
applying common GGA functionals. These functionals assume a fixed gauge of
magnetization direction, i.e. collinear magnetization is always directed on the
$z$-axis. To consistently rotate derivatives when approaching the
zero-magnetization limit, the eigenspace structure has to converge to this
fixed gauge.

Because the choice of
spin-directions becomes a choice of gauge, we regularize $\vec{d}$ so that it
smoothly approaches $(0,0,1)$ in the limit $s=\abs{\vec{d}}=0$ with the regulator function
\begin{equation}
\mathcal{R}(x) = 1 - \mathrm{exp}\parantheses{-x^2/\epsilon^2}
\end{equation}
to obtain the regularized local direction:
\begin{equation}
\overline{\vec{d}} = \mathcal{R}(\abs{\vec{d}}) \vec{d} + \left(1 - \mathcal{R}(\abs{\vec{d}})\right) \begin{pmatrix}0\\0\\1\\\end{pmatrix}\,,
\end{equation}
which guarantees to approach the correct joint collinear limit for vanishing
magnetization.

Numerical diagonalization of $\bm{\gamma}$ will lead to some
non-uniqueness of eigenvectors, when it becomes nearly diagonal,
i.e.  $\frac{\Delta}{\mathrm{max}(\abs{\abs{\bm{\gamma}}}_\mathrm{max},1)}
< \epsilon$, where $\epsilon$ is the numerical precision and the estimate for
the maximal eigenvalue distance is
\begin{equation}
\Delta=\sqrt{\mathrm{max}_\mathrm{off}^2
+ ( \mathrm{max}(\mathrm{diag}(\bm{\gamma})) -
\mathrm{min}(\mathrm{diag}(\bm{\gamma})) )^2}\,,
\end{equation}
with $\mathrm{max}_\mathrm{off}=\abs{\abs{\bm{\gamma}-\mathrm{diag}(\mathrm{diag}(\bm{\gamma}))}}_\mathrm{max}$.

For this case, it is valid to
set $\tilde{\bm{\gamma}}=\mathrm{diag}(\mathrm{diag}(\bm{\gamma}))$, with
$\bm{U}_\gamma=\bm{1}_{4\times4}$. Here $\mathrm{diag}(\bm{M})$ denotes the
vector of diagonal elements of $\bm{M}$ and $\mathrm{diag}(\vec{v})$
denotes a diagonal matrix with the elements of vector $\vec{v}$ occupying its
diagonal. 

In case of non-zero but nearly zero off-diagonal matrix elements,
diagonalization with standard numerical methods can lead to arbitrary
directions in eigenvectors. In order to guarantee a smooth convergence to the
collinear limit along $m_z$, we apply a regularizing level shift 

\begin{equation}
\bm{\lambda} =
\sqrt{\epsilon}\,S_\gamma\brackets{\mathcal{R}\parantheses{\frac{\Delta}{S_\gamma}}-1}\pauli^3\otimes\pauli^3\,,
\end{equation} 
with $S_\gamma=\mathrm{max}(\abs{\abs{\bm{\gamma}}}_\mathrm{max},1)$

The level shift is chosen to quickly vanish for $\Delta>\epsilon$ to
guarantee that no numerical noise is introduced. $\bm{\lambda}$ enforces a
spectral gap that dominates residual couplings, ensuring eigenvector stability
in the intermediate and near-diagonal regimes and a smooth convergence to
$\bm{U}\rightarrow\bm{1}$ in the collinear limit of the XC functional.

We note that the present regularization scheme is not stable if density
matrices contain predominantly numerical noise ($\gtrsim10^{-10}$) without
physically significant spin polarization. This can be the case if, e.g. spin
density matrices are constructed for a closed-shell system. Before applying a
noncollinear method it should be tested if density matrices are significant
$\abs{\abs{\bm{D}^{(\mu)}}}_\mathrm{\max}>\sqrt{\epsilon}$.

In addition, it must be ensured that the eigenspace derivatives become
isotropic when approaching the case of vanishing magnetization. To achieve this
we compute $\gamma_\mathrm{s}=4\Sum{k=1}{3} \gamma_{0k}^2$ and regularize the
eigenspace density operators as
\begin{equation}
\begin{aligned}
\overline{\widetilde{\bm{O}}^{(2)}_{\mu\nu}}&= \mathcal{R}(\gamma_\mathrm{s}) \widetilde{\bm{O}}^{(2)}_{\mu\nu}\\ 
&+\parantheses{1-\mathcal{R}(\gamma_\mathrm{s})} \parantheses{\delta_{0\mu}\delta_{0\nu} \bm{1}_{4\times4} + \parantheses{1-\delta_{0\mu}\delta_{0\nu}}\delta_{\mu\nu}\pauli^3\otimes\pauli^3}\,,\\
\overline{\widetilde{\bm{O}}^{(4)}_{\mu\nu,\kappa\tau}}&= \mathcal{R}(\gamma_\mathrm{s}) \widetilde{\bm{O}}^{(4)}_{\mu\nu,\kappa\tau} \\
&+\parantheses{1-\mathcal{R}(\gamma_\mathrm{s})} \parantheses{ 
\delta_{0\mu}\delta_{0\nu}\delta_{0\kappa}\delta_{0\tau} \bm{1}_{16\times16} \right.\\
&\left.+ \delta_{\mu\nu}\parantheses{1-\delta_{0\nu}}\delta_{\kappa\tau}\parantheses{1-\delta_{0\kappa}}\parantheses{\pauli^3\otimes\pauli^3}^{\otimes2}\right.\\ 
&\left.+ \delta_{\mu0}\delta_{\nu0}\delta_{\kappa\tau}\parantheses{1-\delta_{0\kappa}}\parantheses{\pauli^0\otimes\pauli^0}\otimes\parantheses{\pauli^3\otimes\pauli^3}\right.\\ 
&\left.+ \delta_{\mu\nu}\parantheses{1-\delta_{0\mu}}\delta_{\kappa0}\delta_{\tau0}\parantheses{\pauli^3\otimes\pauli^3}\otimes\parantheses{\pauli^0\otimes\pauli^0}\right.\\ 
&\left.+\delta_{\mu0}\delta_{\kappa0}\delta_{\nu\tau}\parantheses{1-\delta_{0\tau}} \parantheses{\frac{\parantheses{\pauli^0\otimes\pauli^3+\pauli^3\otimes\pauli^0}}{2}}^{\otimes2}\right.\\ 
&\left.+\delta_{\nu0}\delta_{\kappa0}\delta_{\mu\tau}\parantheses{1-\delta_{0\tau}} \parantheses{\frac{\parantheses{\pauli^0\otimes\pauli^3+\pauli^3\otimes\pauli^0}}{2}}^{\otimes2} }
\end{aligned}
\end{equation}

With this regulator magnetization directions, which are distinct for non-zero
$\gamma_{0k}$, become smoothly isotropic for $\gamma_{0k}\rightarrow0$.

\section{\label{sec: numdivdif} Smooth divided differences}
For a divided difference function 
\begin{equation}
\mathcal{D}(a,b) =
\begin{cases}
\frac{\frac{\delta \tilde{F}}{\delta
a}-\frac{\delta
\tilde{F}}{\delta b}}{a-b}
& a \not= b\\
\frac{1}{2}\parantheses{\frac{\delta^2
\tilde{F}}{\delta a^2}+\frac{\delta^2
\tilde{F}}{\delta b^2}} - \frac{\delta^2
\tilde{F}}{\delta a\delta b}  &
a = b\\
\end{cases}\,,
\end{equation}
we define the scale of the numerator $S_\mathrm{num}=\abs{\frac{\delta \tilde{F}}{\delta
a}}+\abs{\frac{\delta
\tilde{F}}{\delta b}}$, the scale of the denominator $S_\mathrm{den}=\mathrm{max}\parantheses{\abs{a},\abs{b}}$, the degeneracy limit as $\Delta_\mathrm{lim}=\frac{1}{2}\parantheses{\frac{\delta^2
\tilde{F}}{\delta a^2}+\frac{\delta^2
\tilde{F}}{\delta b^2}} - \frac{\delta^2
\tilde{F}}{\delta a\delta b}$, as well as the denominator $\Delta_\mathrm{den}=a-b$ and numerator $\Delta_\mathrm{num}=\frac{\delta \tilde{F}}{\delta
a}-\frac{\delta
\tilde{F}}{\delta b}$. A smooth switching function, which is defined in the range $[0,1]$, can be computed as:
\begin{equation}
\mathcal{S} = \mathrm{max}\parantheses{0,\mathrm{min}\parantheses{1,\mathrm{min}\parantheses{\frac{\Delta_\mathrm{num}}{S_\mathrm{num}\sqrt{\epsilon}},\frac{\Delta_\mathrm{den}}{S_\mathrm{den}\sqrt{\epsilon}}}}}\,,
\end{equation}
This function was chosen because it provides a bounded, scale-aware measure of
the reliability of the numerator and denominator in the divided difference. It
suppresses contributions when values fall below the floating-point noise level
and thereby improves numerical stability while avoiding cancellation-induced
amplification.  With $\mathcal{S}$ we can define a smooth divided difference
function as
\begin{equation}
\overline{\mathcal{D}(a,b)} = \mathcal{S} \frac{\Delta_\mathrm{num}}{\Delta_\mathrm{den}} + (1-\mathcal{S})\Delta_\mathrm{lim}\,.
\end{equation}

\section{Magnetic hyperfine coupling Hamiltonian \label{sec: hyperfine}}
The nuclear magnetization density distribution was assumed to coincide with the
nuclear charge densities. The hyperfine
coupling operator appears in this approximation as  
\begin{equation}
\hat{H}_{\mathrm{hf}} = \frac{\mu_0}{4\pi}\Sum{i}{N}c\, e\, \diraca_i\cdot\vec{\mathcal{A}}(\pos_i)
\end{equation}
where the vector potential of a spherical Gaussian nuclear magnetization distribution is
\begin{equation}
\vec{\mathcal{A}}(\pos_i) = 4\sqrt{\frac{\zeta^3}{\pi}} \hat{I}(\ce{^{233}U})\times(\pos_i-\pos_{\mathrm{U}}) F_1(\zeta \abs{\pos_i-\pos_{\mathrm{U}}}^2)\,,
\end{equation}
where $F_n(x)=\int_0^1\mathrm{d}t\mathrm{e}^{-xt^2}t^{2n}$ is the Boys
function.  The Dirac matrix is
$\diraca=\begin{pmatrix}\bm{0}&\vec{\pauli}\\\vec{\pauli}&\bm{0}\end{pmatrix}$.
The transformation of the operator to the 2c-ZORA picture follows
Refs.\,\cite{gaul:2020,gaul:2020a}.

\bibliography{AK}

\end{document}